\documentclass[12pt]{article}
\usepackage{txfonts}
\usepackage{amssymb}

\usepackage{scicite}

\usepackage{CJK}
\usepackage{subfigure}
\usepackage{times}
\usepackage{epsfig}
\usepackage{url}

\newenvironment{sciabstract}{%
\begin{quote} \bf}
{\end{quote}}

\newcounter{lastnote}

\title{ Homophyly and Randomness Resist Cascading Failure in Networks}

\author{ Angsheng Li$^{1}$, Wei Zhang$^{1,2}$\and Yicheng Pan$^{1,3}$,  Xuechen Li$^{4}$ \\
\normalsize{$^{1}$State Key Laboratory of Computer Science}\\
\normalsize{ Institute of Software, Chinese Academy of Sciences, P. R. China}\\
\normalsize{$^{2}$University of Chinese Academy of
Sciences, P. R. China}\\
\normalsize{$^{3}$State Key Laboratory of Information Security}\\
\normalsize{ Institute of Information Engineering, Chinese Academy of Sciences, P. R. China}\\
\normalsize{$^{4}$Beijing No. 4 High School, P. R. China} }

\date{}

\begin{document}


\baselineskip24pt


\maketitle

\begin{sciabstract}

The universal properties of power law and small world phenomenon of
networks seem unavoidably obstacles for security of networking
systems. Existing models never give secure networks. We found that
the essence of security is the security against cascading failures
of attacks and that nature solves the security by mechanisms. We
proposed a model of networks by the natural mechanisms of homophyly,
randomness and preferential attachment. It was shown that homophyly
creates a community structure, that homophyly and randomness
introduce ordering in the networks, and that homophyly creates
inclusiveness and introduces rules of infections. These principles
allow us to provably guarantee the security of the networks against
any attacks. Our results show that security can be achieved provably
by structures, that there is a tradeoff between the roles of
structures and of thresholds in security engineering, and that power
law and small world property are never obstacles for security of
networks.

\end{sciabstract}

Network security has become a grand challenge in the
current science and technology. We proposed a mathematical
definition of network security, and a new model of networks by
natural mechanisms of homophyly, randomness and preferential
attachment. We found that networks of our model satisfy a serious of
new topological, probabilistic and combinatorial principles, and
that the new principles ensure that the networks are provably
secure. Our model provides a foundation for both theoretical and
practical analyses of security of networks. Generally, our model
demonstrates that nature may solve security of complex systems by
mechanisms, exploring a new principle for networking systems in
nature, society, economics, industry and technology etc.

Many real networks satisfy the power law ~\cite{Ba1999, Ba2009,
CL2006}, and the small world phenomenon~\cite{WS1998, JK2000,
W2002}. A surprising discovery in network theory in the first $10$
years after the discovery of power law in \cite{Ba1999}, is perhaps
that network topology is universal in nature, society and industry
\cite{Ba2009}. This universality allowed researchers from different
disciplines to embrace network theory as a common paradigm. The
understanding of networks is a common goal of an unprecedented array
of traditional disciplines: For instance, cell biologists use
networks to capture signal transduction cascade and metabolism;
computer scientists are mapping the Internet and the WWW;
epidemiologists follow transmission networks trough which virus
spread \cite{Ba2009}.

From the second decade of network theory, security of networks has
become a sharper focus and a grand challenge. We have to understand
how the internet responds to attacks and traffic jams, or how the
cell reacts to changes in its environments, or how the global
economy responses to the current financial crisis, or even how a
society reacts to a social crisis. A basic question of this issue is
the security of networks.

To understand the essence of security of networks, we examine the
two classic models of networks. The first is the Erd\"os-R\'enyi
model~\cite{ER1959, ER1960}. In this model, we are given $n$ nodes,
and a number $p$, and create an edge with probability $p$ for each
pair of nodes. The second is the preferential attachment (PA, for
short) model~\cite{Ba1999}. In this model, for a given initial
graph, $G_0$ say, and a natural number $d$, we build the network $G$
by steps. Suppose that $G_{t-1}$ is defined. At step $t$,  we create
a new node, linking to $d$ nodes chosen with probability
proportional to the degrees of nodes in $G_{t-1}$.

Security must depend on strategies of attacks. Typical strategies
are the physical attack of removal of nodes or edges, and the
cascading failures of attacks.

In~\cite{AJB2000, CRB2000, M2004}, it has been shown that in
scale-free networks of the PA model, the overall network
connectivity measured by the sizes of the giant connected components
and the diameters does not change significantly under random removal
of a small fraction of nodes, but is vulnerable to removal of a
small fraction of the high degree nodes.

In~\cite{N2003,W2002,AM1991,M2000}, the cascading failure model was
proposed to study rumor spreading, disease spreading, voting, and
advertising etc. In~\cite{PV2001}, it has been shown that in
scale-free networks of the PA model even weakly virulent virus can
spread.

{\bf The Essence of Network Security}

Let $G=(V,E)$ be a network. Suppose that for each node $v\in V$,
there is a threshold $\phi (v)$ associated with it. For an initial
set $S\subset V$, the {\it infection set} of $S$ in $G$ is defined
recursively as follows: (1) Each node $x\in S$ is called {\it
infected}, and (2) A node $x\in V$ becomes infected, if it has not
been infected yet, and $\phi (x)$ fraction of its neighbors have
been infected. We use ${\rm inf}_G(S)$ to denote the infection set
of $S$ in $G$.

The cascading failure models depend on the choices of thresholds
$\phi (v)$ for all $v$. We consider two natural choices of the
thresholds. The first is random threshold cascading, and the second
is uniform threshold cascading. We say that a cascading failure
model is {\it random}, if for each node $v$, $\phi (v)$ is defined
randomly and uniformly, that is, $\phi (v)=r/d$, where $d$ is the
degree of $v$ in $G$, and $r$ is chosen randomly and uniformly from
$\{1,2,\cdots, d\}$. We say that a cascading failure model is {\it
uniform}, if for each node $v$, $\phi (v)=\phi$ for some fixed
number $\phi$.

To understand the nature and essence of security of networks, we
compare the two strategies of physical attacks and the cascading
failure models of attacks. For this, we introduce the notion of {\it
injury set} of physical attacks. Let $G=(V,E)$ be a network, and $S$
be a subset of $V$. The physical attacks on $S$ is to delete all
nodes in $S$ from $G$. We say that a node $v$ is injured by the
physical attacks on $S$, if $v$ is not connected to the largest
connected component of the graph obtained from $G$ by deleting all
nodes in $S$. We use ${\rm inj}_G(S)$ to denote the injury set of
$S$ in $G$.

\begin{figure}
  \centering
  \subfigure[]
   {\label{fig:_cascading_vs_node_attack_ER_N=10000_d=10}
    \includegraphics[width=4in]{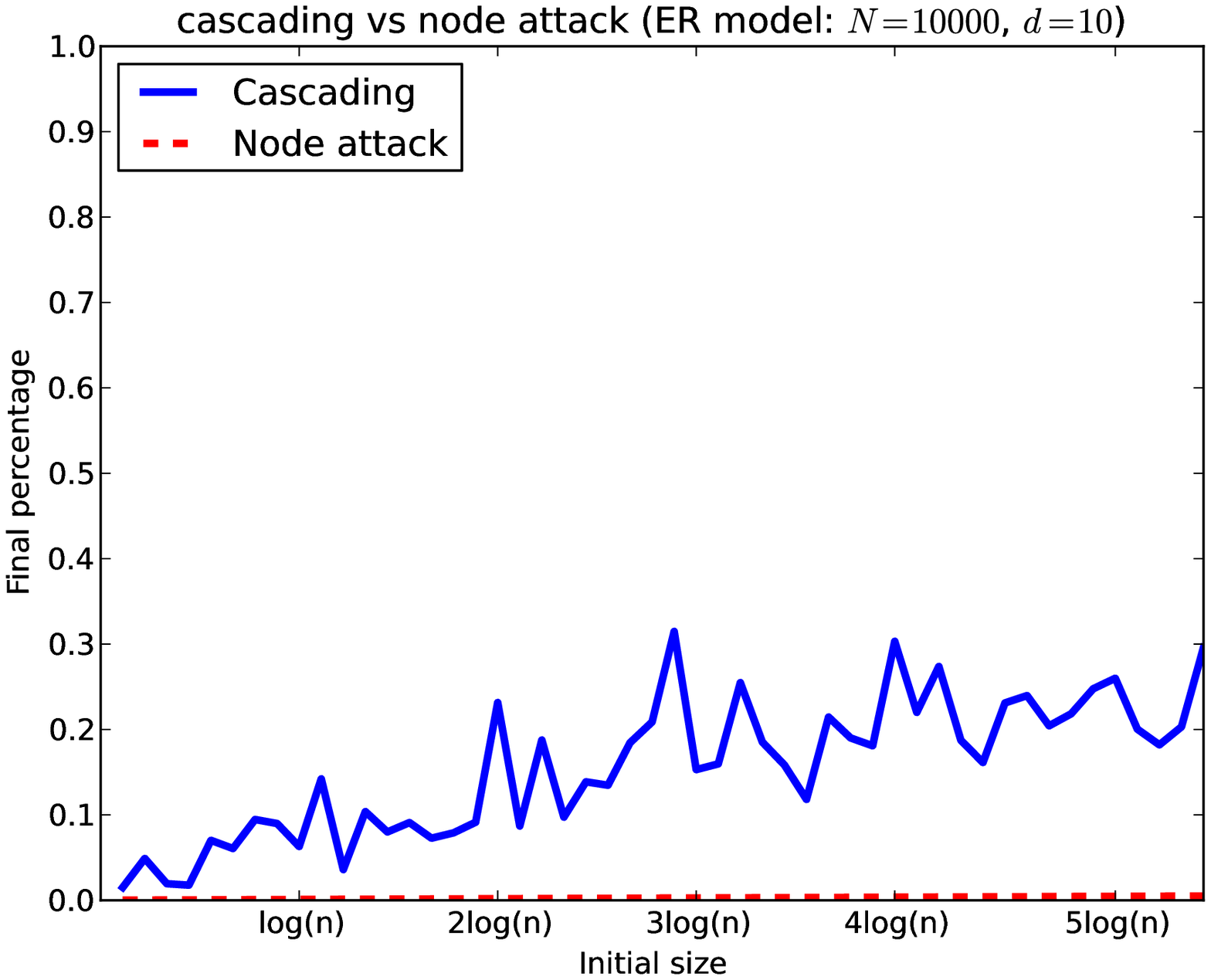}
   }

   \subfigure[]
   {\label{fig:_cascading_vs_node_attack_PA_N=10000_d=10}
    \includegraphics[width=4in]{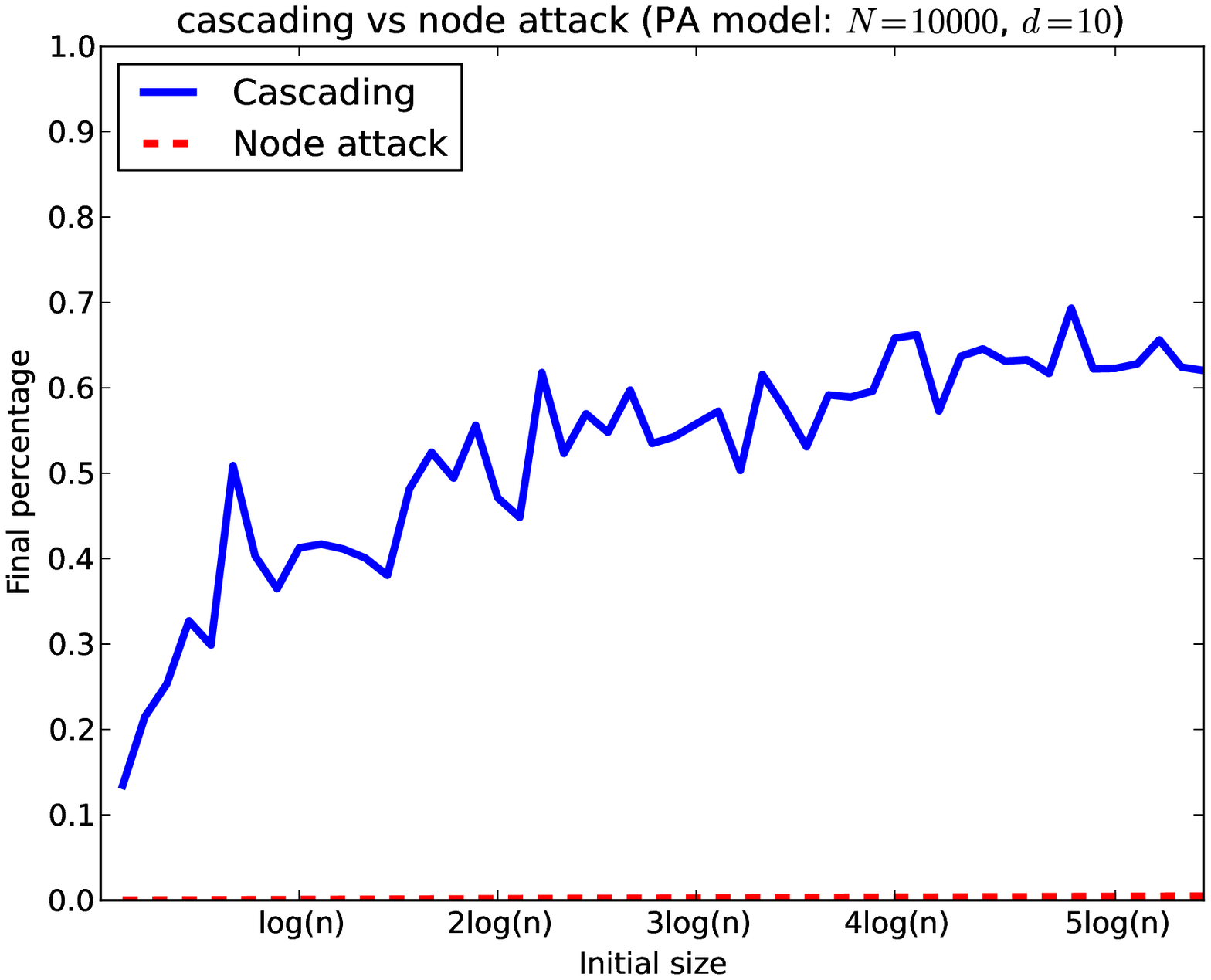}
   }

  \caption{\textbf{(a), (b) are the curves of cascading failure and injured nodes by nodes
   removal of networks for $n=10,000$ and $d=10$  of the ER and PA models respectively. The red curves are
    fractions of injury sets of attacks on the top degree nodes of size up to $5\cdot\log n$, and the curves
     colored blue are the fractions of the largest infection sets
    among $100$ times of attacks of the top degree nodes of size less than $5\cdot\log n$ under the random threshold cascading failure model.}}
\end{figure}

We depict the curves of sizes of the infection sets and the injury
sets of attacks of top degree nodes of networks of the ER and PA
models in Figure
~\ref{fig:_cascading_vs_node_attack_ER_N=10000_d=10}, and
Figure~\ref{fig:_cascading_vs_node_attack_PA_N=10000_d=10}
respectively. From the figures, we know that for any network, $G$
say, generated from either the ER or the PA model, the following
properties hold: (1) the infection sets are much larger than the
injury sets, (2) the attacks of top degree nodes of size as small as
$O(\log n)$ may cause a constant fraction of nodes of the network to
be infected under the cascading failure models of attacks, and (3)
structures play a role in security of networks, by observing the
difference between Figure
~\ref{fig:_cascading_vs_node_attack_ER_N=10000_d=10}, and
Figure~\ref{fig:_cascading_vs_node_attack_PA_N=10000_d=10}.

The experiments in Figures
~\ref{fig:_cascading_vs_node_attack_ER_N=10000_d=10} and
\ref{fig:_cascading_vs_node_attack_PA_N=10000_d=10} show that the
essence of network security is the security against any attacks of
sizes polynomial in $\log n$ under the cascading failure models.

Let $\mathcal{M}$ be a model of networks. We investigate the
security of networks of model $\mathcal{M}$. We define the security
of networks under both uniform and random threshold cascading
failure models.

Let $G$ be a network of $n$ nodes constructed from model
$\mathcal{M}$. For the random threshold cascading failure model, we
say that $G$ is {\it secure}, if almost surely (meaning that with
probability arbitrarily close to $1$ as $n$ grows), the following
holds: for any set $S$ of size bounded by a polynomial of $\log n$,
the size of the infection set of $S$ in $G$ is bounded by $o(n)$,
meaning that it is negligible comparing with $n$. For the uniform
threshold cascading failure model, we say that $G$ is {\it secure},
if almost surely, the following holds:  for some arbitrarily small
$\phi$, i.e., $\phi=o(1)$, for any set $S$ of size bounded by a
polynomial of $\log n$, the infection set of $S$ in $G$ with uniform
threshold $\phi$ has size $o(n)$.

{\bf Questions and Results}

By the definitions of security of networks, and by the experiments
in Figures ~\ref{fig:_cascading_vs_node_attack_ER_N=10000_d=10}, and
\ref{fig:_cascading_vs_node_attack_PA_N=10000_d=10}, we have that
both the ER and the PA models never give secure networks.

Notice that randomness is the mechanism of the ER model, and is in
fact the mechanism for the small world property for almost all
networks (by observing all other models and real networks), and that
preferential attachment is the mechanism of the PA model which
guarantees the power law of the networks. The experiments in Figures
~\ref{fig:_cascading_vs_node_attack_ER_N=10000_d=10}, and
\ref{fig:_cascading_vs_node_attack_PA_N=10000_d=10} show that
neither randomness nor preferential attachment alone is a mechanism
for security of networks. This also implies that small world
property and power law seem obstacles for security of networks.

The fundamental questions are thus: Are power law and small world
property really obstacles for security of networks? What mechanisms
and principles can guarantee security of networks? Is there an
algorithm to construct secure networks?  In this paper, we will
answer these questions.

We found that homophyly is a new mechanism of networks, that
homophyly guarantees a community structure of networks, that
homophyly and randomness introduce ordering in networks and generate
a degree priority principle, that homophyly creates inclusiveness
and introduces infection rules in networks. These discoveries allow
us to give an algorithm based on natural mechanisms of homophyly,
randomness or uncertainty and preferential attachment to construct
networks such that the networks are provably secure, follow a power
law, have the small diameter property, and furthermore, have a
navigation algorithm of time complexity $O(\log n)$.

The results show that security can be achieved by structures of
networks, that there exists a tradeoff between the role of structure
and the role of thresholds in security of networks, and that neither
power law nor small world property is an obstacle of security of
networks.

{\bf  Security Model}

How can we construct secure networks? Networks are proved universal
in a wide range of disciplines in both nature and society. This
suggests that natural mechanisms of the evolution of complex systems
in nature and society maybe helpful for us to construct secure
networks.

Let us consider a mental experiment in evolution of networking
systems in nature. Assume that $H$ is the current network. Suppose
that a new individual $v$ is born. Then $v$ has its own
characteristic from the very beginning of its birth either as a
remarkable element or a normal element. If $v$ is born as a
remarkable element, then it develops some links to individuals in
$H$ by the preferential attachment scheme in the whole $H$, and some
links to remarkable elements in $H$ by chance, and $v$ will develop
its own community. If $v$ is born as a normal individual, then it is
very likely that $v$ joins randomly some group of individuals, in
which case, $v$ links to some individuals in that group by a
preferential attachment scheme.

Based on this mental experiment, we propose a new model of networks,
the {\it security model} below.

The security model proceeds as follows: (1) Given a {\it homophyly
exponent $a$} and a natural number $d$, let $G_d$ be an initial
$d$-regular graph. Each node of $G_d$ is associated with a distinct
color and called a {\it seed}. For $i>d$, let $G_{i-1}$ be the graph
constructed at the end of step $i-1$. At step $i$, set $p_i=1/(\log
i)^a$. (2) At time step $i$, create a new node $v$. (3) With
probability $p_i$, $v$ chooses a new color, $c$ say, in which case:
(a) we say that $v$ is a seed node, (b) (PA scheme) add one edge
$(v,u)$ such that $u$ is chosen with probability proportional to the
degrees among all nodes in $G_{i-1}$, and (c) (Randomness) add $d-1$
edges $(v,u_j)$ for $j=1,2,\cdots, d-1$, where $u_j$ is chosen
randomly and uniformly among all seed nodes in $G_{i-1}$. (4)
Otherwise, then $v$ chooses an old color, in which case: (a)
(Randomness) $v$ chooses randomly and uniformly an old color, and
(b) (Homophyly and PA scheme) create $d$ edges from $v$ to nodes of
the same color as $v$  chosen with probability proportional to the
degrees of the nodes in $G_{i-1}$.

Obviously the model is dynamic. The mechanisms of the model are
homophyly, randomness (or uncertainty) and preferential attachment.
Clearly, each of the three mechanisms is a natural mechanism in
evolution of networking systems in nature and society.

{\bf Mathematical Principles}

We will show that networks generated from the security model are
secure against any attacks of small-scales under both uniform and
random threshold cascading failure models.

The authors have shown that networks of the
security model satisfy four groups of topological, probabilistic and
combinatorial principles (A. Li, Y. Pan and W. Zhang, Provable security of networks).

Let $a>1$ be the homophyly exponent, and $d\geq 4$ be a natural
number. Let $G=(V,E)$ be a network constructed by our model. Then
with probability $1-o(1)$, $G$ satisfies the following four
principles each of which consists of a number of interesting
properties.

The first is a {\it fundamental principle}, consisting of a number
of topological and probabilistic properties: (1) (Basic properties):
(i) The number of seed nodes is bounded in the interval
$[\frac{n}{2\log^a n},\frac{2n}{\log^a n}]$, and (ii) Each
homochromatic set has a size bounded by $O(\log^{a+1} n)$; (2) For
degree distributions, we have: (i) The degrees of the induced
subgraph of a homochromatic set follow a power law, (ii) The degrees
of nodes of a homochromatic set follow a power law, and (iii) (Power
law) Degrees of nodes in $V$ follow a power law; (3) For
node-to-node distances, we have: (i) The induced subgraph of a
homochromatic set has a diameter bounded by $O(\log\log n)$, (ii)
(Small world phenomenon) The average node to node distance of $G$ is
bounded by $O(\log n)$, and (iii) (Local algorithm for navigating)
There is an algorithm to find a short path between arbitrarily given
two nodes in time complexity $O(\log n)$; and (4) (Small community
phenomenon) There are $1-o(1)$ fraction of nodes of $G$ each of
which belongs to a homochromatic set, $W$ say, such that the
conductance of $W$, $\Phi (W)$, is bounded by
$O\left(\frac{1}{|W|^{\beta}}\right)$ for
$\beta=\frac{a-1}{4(a+1)}$.

We define a community of $G$ to be the induced subgraph of a
homochromatic set. By the fundamental principle, we know that all
the communities are small, that $G$ has both a power law local
structure and a power law global structure, that $G$ has not only a
short diameter, but also a local algorithm of time complexity
$O(\log n)$ to find a short path between arbitrarily given two
nodes, and that $G$ has a remarkable community structure.

The second is a {\it degree priority principle}, consisting of some
properties of the degree priority of vertices of $G$.

Let $v$ be a node of $G$. We consider the homochromatic sets of all
the neighbors of $v$. We define the {\it length of degrees of $v$}
to be the number of colors of the neighbors of $v$, written by
$l(v)$. For each $j\in\{1,2,\cdots,l(v)\}$, let $X_j$ be the $j$-th
largest homochromatic set of all the neighbors of $v$ (break ties
arbitrarily). We define the {\it $j$-th degree of $v$} to be the
size of $X_j$.

Then we have the following degree priority principle:

 For
a randomly chosen node $v$, with probability $1-o(1)$, the following
properties hold: (1) The length of degrees of $v$ is bounded by
$O(\log n)$, (2) The first degree of $v$ is the number of $v$'s
neighbors that share the same color as $v$, (3) The second degree of
$v$ is bounded by $O(1)$, so that for any possible $j>1$, the $j$-th
degree of $v$ is $O(1)$, and (4) The first degree of a seed node is
at least $\Omega (\log^{\frac{a+1}{4}}n)$.

The third one is an {\it infection-inclusion principle}, created by
homophyly and randomness of the model.

 Let $x$ and $y$ be two nodes of $G$. We say that $x$
injures $y$, if the infection of $x$ contributes to the probability
that $y$ becomes infected. Otherwise, we say that $x$ fails to
injure $y$.

 Let $X$ and $Y$ be
 two homochromatic sets. Suppose that $G_X$ and $G_Y$
are the induced communities by $X$ and $Y$ respectively. Let $x_0$,
and $y_0$ be the seed nodes of $X$, and $Y$ respectively. Suppose
that $x_0$ and $y_0$ are created at time step $s$ and $t$
respectively. The infection-inclusion principle ensures that with
probability $1-o(1)$, the following properties hold: (1) If $s<t$,
then: (i) community $G_X$ fails to injure any non-seed node in
community $G_Y$, and (ii) the number of neighbors of the seed node
$y_0$ that are in $X$ is bounded by a constant $O(1)$; and (2) If
$s>t$, then: (i) all the non-seed nodes in $G_X$ fail to injure any
node in community $G_Y$, (ii) the number of neighbors of the seed
node $y_0$ that are in $X$ is bounded by $1$, and (iii) the injury
of a non-seed node in $Y$ from the seed node of $X$ follows only the
edge created by step (3) (b) of the definition of the model.

The infection-inclusion principle shows that homophyly creates some
inclusiveness among the non-seed nodes, and that a community
protects its non-seed members from being arbitrarily injured by the
collection of their neighbor communities.

The fourth one is an {\it infection priority tree principle}. We
define the infection priority tree $T$ of $G$ as follows: (i) for
each edge $e=(u,v)$ in $G$, if $u$ and $v$ were created at time
steps $s>t$ respectively, then we interpret the edge $e=(u,v)$ as a
directed edge from $u$ to $v$, (ii) let $H$ be the graph obtained
from $G$ by deleting all edges created by (3) (c) of definition of
our model, and (iii) let $T$ be the graph obtained from $H$ by
merging each of the homochromatic set into a single node, and at the
same time, keeping all the directed edges.

We have the following infection priority tree principle: (1) $T$ is
a tree on which the injury directions always going to the early
created nodes, and (2) with probability $1-o(1)$, the infection
priority tree $T$ has a height bounded by $O(\log n)$.

Note that the direction in $T$ is determined by the injury of a
non-seed node from a seed node of a neighbor community as shown in
the infection-inclusion principle.

{\bf Proofs of Security }

By combining the four principles together, we are able to prove some
security results.

Let $G$ be a network constructed by our model. By the fundamental
principle, all the communities are small, and the number
 of seed nodes is large. Let $X$ be a homochromatic set with seed $x_0$. Suppose that there
 is no node in $X$ which has been targeted.
 By the degree priority principle, the first and second degrees of $x_0$ is at least
$\Omega (\log^{\frac{a+1}{4}}n)$, and at moat $O(1)$ respectively.
Therefore, the seed node $x_0$ of $G_X$ is hard to be infected by a
single neighbor community $G_Y$, if any. Furthermore, by the same
principle, the length of degrees of $x_0$ is at most $O(\log n)$,
therefore, for properly chosen $a$, the seed node $x_0$ of $G_X$ is
hard to be infected by the collection of all its neighbor
communities.

We say that a community $G_X$ is {\it strong}, if the seed $x_0$ of
$X$ can not be infected by the collection of all its neighbor
communities alone, and {\it vulnerable}, otherwise. The degree
priority principle ensures that for properly chosen $a$, almost all
communities are strong.

By definition of the infection priority tree $T$, and by the
infection-inclusion principle, infections among strong communities
from its neighbor communities must be triggered by an edge in the
infection priority tree $T$ of $G$. We further explain this as
follows.

Suppose that $G_X$, $G_Y$ and $G_Z$ are strong communities. Let
$x_0$, $y_0$ and $z_0$ be the seed nodes of $X$, $Y$ and $Z$
respectively. Suppose that $x_0$, $y_0$ and $z_0$ are created at
time step $t_1$, $t_2$ and $t_3$ respectively.

Then it is possible that $x_0$ infects a non-seed node $y_1\in Y$,
$y_1$ infects the seed node $y_0$ of $Y$ and $y_0$ infects a
non-seed node $z_1\in Z$. By the infection-inclusion principle, we
have that $t_1>t_2>t_3$, and that the edges $(x_0, y_1)$ and $(y_0,
z_1)$ are created by step (3) (b) of the construction of the network
so that the edges are embedded in the infection priority tree $T$ of
$G$.

By the {\it infection priority tree principle}, $T$ is directed with
direction always going towards the early created nodes, and $T$ has
height bounded by $O(\log n)$, with probability $1-o(1)$. Therefore
whenever a strong community triggers an infection in the priority
tree, it generates at most $O(\log n)$ many strong communities to be
infected, where a community is infected, if at least one node of the
community has been infected.

By the fundamental principle, each community has size at most
$O(\log^{a+1}n)$. Therefore an infected community contributes at
most $O(\log^{a+1}n)$ many infected nodes.

By using the ideas above, we estimate the number of infected nodes
by an attack of small scales. By the degree priority principle, for
properly chosen $a$, almost all communities are strong. Let $k$ be
the number of vulnerable communities. Then $k$ must be negligible.

Suppose that $S$ is a set of nodes of size $\log^c n$ for some
constant $c$. We attack all the nodes in $S$. Then there are at most
$|S|+k$ communities each of which triggers an infection in the
infection priority tree $T$. By the infection priority tree
principle, the total number of infected communities is at most
$O((|S|+k)\cdot\log n)$. Therefore even if all the nodes of an
infected community are infected, the total number of nodes that are
infected by attacks on $S$ is at most $O((|S|+k)\log^{a+2}n)$, which
could be $o(n)$.

Now the only problem is to estimate the number of vulnerable
communities $k$ which is some probabilistic arguments. In fact, we have shown that: (1) For
the uniform threshold cascading model, for $a>4$, and $d\geq 4$, let
$G$ be a network constructed by our security model, then with
probability $1-o(1)$, the following event occurs: For some
$\phi=o(1)$, for any constant $c$, and any set $S$ of vertices of
$G$, if $S$ has size bounded by $\log ^cn$, then the infection set
of $S$ in $G$ with uniform threshold $\phi$ has size $o(n)$, where
$n$ is the number of vertices of $G$; and (2) For the random
threshold cascading model, for $a>6$, and
 $d\geq 4$, let $G$ be a network of the security model, then with
 probability $1-o(1)$, the following event occurs: For any constant $c$,
 any set $S$ of vertices of $G$, if the size
 of $S$ is bounded by $\log ^cn$, then the infection set of $S$ in
 $G$ has size $o(n)$.

{\bf Experiments }

Our theoretical results require $a>4$ and $a>6$ for the uniform and
random threshold cascading models respectively. The reason is the
degree priority principle. We know that the lower bound of the first
degree of a seed node is $\Omega (\log^{\frac{a+1}{4}}n)$, which
depends on $a$, and that in the worst case, the degree of the seed
node contributed by all its neighbor communities is $O(\log n)$. To
make sure that a community is strong in the case that the threshold
of the seed node is sufficiently small, we have to choose $a$ to be
appropriately large. Therefore, if $a$ is too small, then the number
of strong communities will be relatively small, in which case, the
network of the model will be less secure. However we will show that
networks of the security model with small $a$ have much better
security than that of the other models.

\begin{figure}[ht]

                 \centering
                 \includegraphics[width=3.2in]{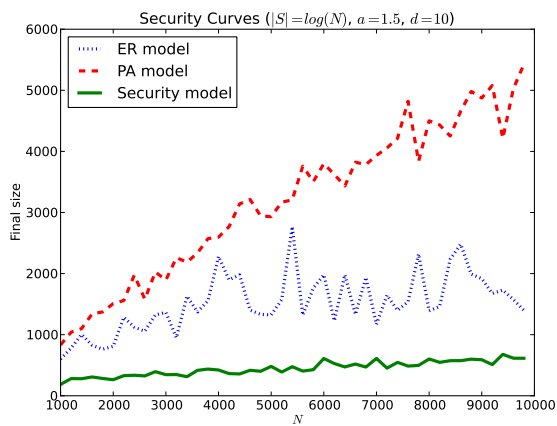}
                 \caption{Security curves of networks of the Erd\"os-R\'enyi model,
                 the preferential attachment model and the security model. In this figure, we consider the
case of random thresholds. It depicts the curves of the greatest
size of the cascading failure sets of attacks of $100$ times of size
 $\log n$ for each $n$ less than or equal to $10,000$. The curves describe the greatest sizes of the final cascading
                 failure sets among $100$ times attacks of the random thresholds. The sizes of the initial attacks are always
                 $\log n$, where $n$ is the number of nodes of the networks.
}
                 \label{fig:Security_plot_d=10_a=1.5}

\end{figure}

\begin{figure}[ht]

                 \centering
                 \includegraphics[width=3.2in]{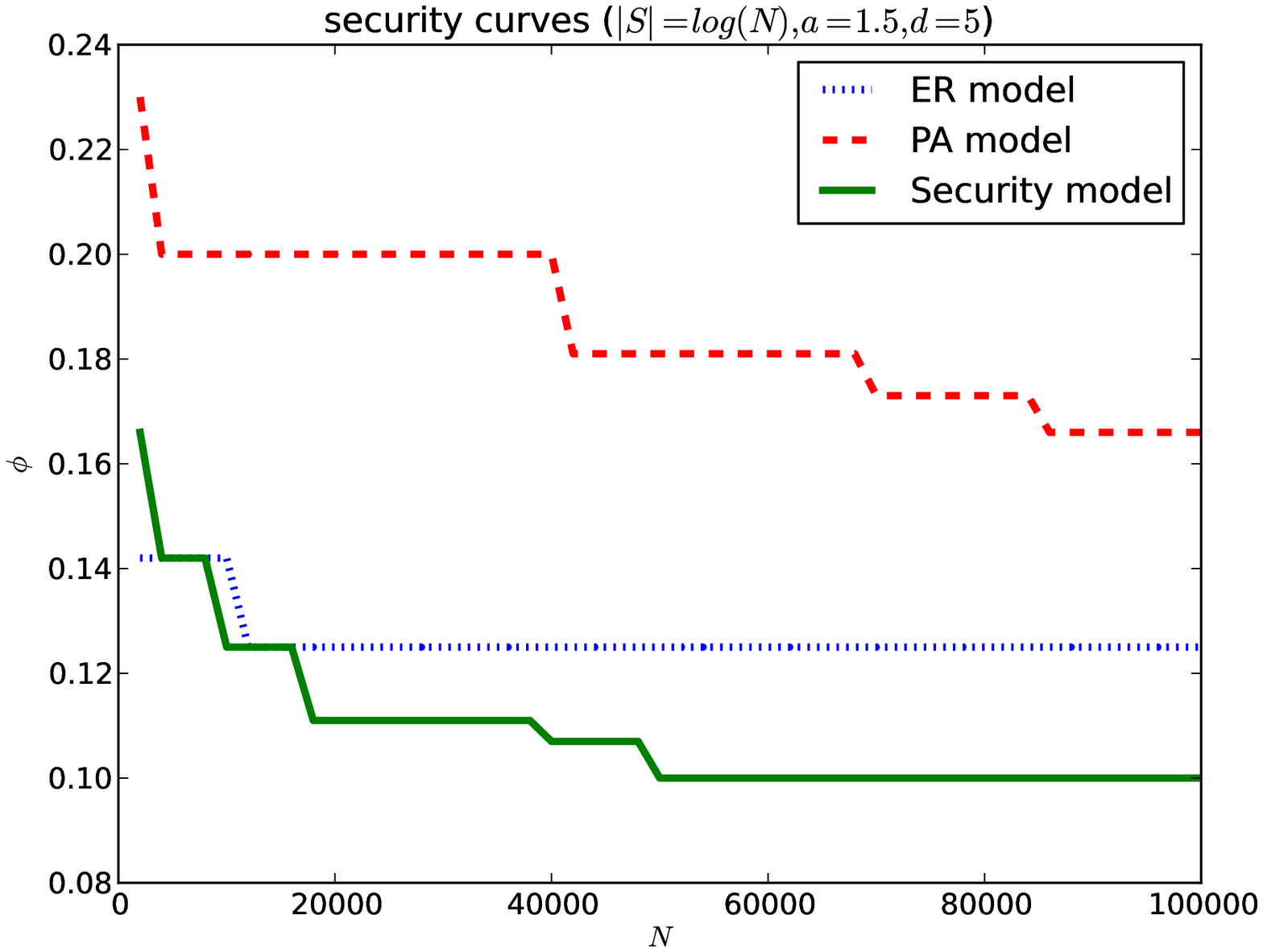}
                 \caption{Security curves for initial size $\log n$, $a=1.5$ and $d=5$. We consider
networks of nodes up to $100,000$, each of which has average number
of edges $d=5$. For the security model, we set the homophyly
exponent $a=1.5$ for all the networks. This describes the curves of
security thresholds of the networks generated from the $3$ models of
nodes up to $100,000$ and of average number of edges $d$ for $d=5$
respectively. In this experiment, the initial set of attacks are the
top degree nodes of size $\log n$. The homophyly exponent $a$ is
chosen as $1.5$ for  the security model.}
                 \label{fig:Phi_security_c=1_a=1.5_d=5.txt}

\end{figure}

Our experiments below show that even if just for $a>1$, the networks
of our model are much more secure than that of the ER and PA models.
From Figure~\ref{fig:Security_plot_d=10_a=1.5}, we have that for any
size $n$, for a network of either the PA model or the ER model, the
attacks of size only $\log n$ would probably generate a global
cascading failure of the network. In sharp contrast to this, for any
$n$, any attacks of $\log n$ size are unlikely to generate a global
cascading failure of the networks generated from the security model.
For the uniform threshold cascading failure model, we compare the
security thresholds of networks of the three models. From
Figure~\ref{fig:Phi_security_c=1_a=1.5_d=5.txt}, we have that the
curve of the security thresholds of networks of the security model
is the lowest, much better than that of the ER and PA models.

In summary, both theoretical analysis and experiments show that
networks of our model resist cascading failure of attacks, for which
homophyly, randomness and preferential attachment are the underlying
mechanisms. Our fundamental principle also shows that networks of
the security model follow a power law and satisfy the small world
property, and more importantly, allowing a navigation algorithm of
time complexity $O(\log n)$. This shows that power law and small
world property are never obstacles of security of networks.

\newpage

\newpage


\begin{thebibliography}{10}

\bibitem{Ba1999}
Barab{\'a}si, A. L. $\&$ Albert, R.
\newblock Emergence of scaling in random networks, 
\newblock {\em Science}, 286, 509--512, (1999).

\bibitem{Ba2009}
Barab{\'a}si, A. L.
\newblock Scale-free networks: a decade and beyond,
\newblock {\em Science}, 325, 412--413, (2009).

\bibitem{CL2006}
Chung, F. $\&$ Lu, L.
\newblock {\em Complex Graphs and Networks},
\newblock (American Mathematical Society, 2006).

\bibitem{WS1998}
Watta, D. J. $\&$ Strogatz, S. H.
\newblock Collective dynamics of small world networks,
\newblock {\em Nature}, 393, 440--442, (1998).

\bibitem{JK2000}
Kleinberg, J.
\newblock Navigation in a small world,
\newblock {\em Nature}, 406, 845, (2000).

\bibitem{W2002}
Watta, D. J.
\newblock A simple model of global cascades on random networks,
\newblock {\em Proceedings of the National Academy of Sciences},
  99, 5766--5771, (2002).

\bibitem{ER1959}
Erd\"os, P. $\&$ R\'enyi, A.
\newblock On random graphs, i,
\newblock {\em Publ. Math.}, 6, 290--297, (1959).

\bibitem{ER1960}
Erd\"os, P. $\&$  R\'enyi, A.
\newblock On the evolution of random graphs, i, 
\newblock {\em Magyar Tud. Akad. Mat. Kutat\'o Int. K\'ozl.}, 5, 17--61, (1960).

\bibitem{AJB2000}
Albert, R.,  Jeong, H. $\&$  Barab{\'a}si, A. L.
\newblock Error and attack tolerance of complex networks,
\newblock {\em Nature}, 406, 378--381, (2000).

\bibitem{CRB2000}
Cohen, R., Erez, K., Ben-Avraham, D. $\&$  Havlin, S.
\newblock Resilience of the internet to random breakdowns, 
\newblock {\em Physical Review Letters}, 85, 4626--4628, (2000).

\bibitem{M2004}
 Motter, A. E.
\newblock Cascade control and defense in complex networks,
\newblock {\em Physical Review Letters}, 93, 098701, (2004).

\bibitem{N2003}
Newman, M. E. J.
\newblock The structure and function of complex networks,
\newblock {\em SIAM Rev.}, 45, 167--256, (2003).


\bibitem{AM1991}
Andersen, R.~M. $\&$ May, R.~M.
\newblock {\em Infectious diseases of humans: Dynamics and control}.
\newblock { (Oxford University Press, 1991)}.

\bibitem{M2000}
Morris, S.
\newblock Contagion.
\newblock {\em Review of Economic Studies}{ \bf 67}, 57--78 (2000).

\bibitem{PV2001}
Pastor-Satorras, R.,  V{\'a}zquez, A. $\&$ Vespignani, A.
\newblock Dynamical and correlation properties of the internet,
\newblock {\em Physical Review Letters}, 87, 258701, (2001).



\end{thebibliography}

{\bf Acknowledgements}

Angsheng Li is partially supported by the Hundred-Talent Program of
the Chinese Academy of Sciences. He gratefully acknowledges the
support of the Isaac Newton Institute for Mathematical Sciences,
Cambridge University, where he was a visiting fellow during the
preparation of this paper. All authors are partially supported by
the Grand Project ``Network Algorithms and Digital Information'' of
the Institute of Software, Chinese Academy of Sciences, and by an
NSFC grant No. 61161130530 and a
973 program grant No. 2014CB340302. Yicheng Pan is partially supported by a National Key Basic Research Project of China (2011CB302400)
and  the "Strategic Priority Research Program" of the Chinese Academy of
Sciences£¬Grant No. XDA06010701. Xuechen Li is partially supported by a
program of China Science Future Star.

{\bf Author Contributions} AL designed the research and wrote the
paper, WZ, YP and XL performed the research. All authors reviewed the paper.

{\bf Additional information }

Competing financial interests: The authors declare they have no
competing financial interests.

\end{document}